\documentclass[preprint,10pt]{aastex}
\usepackage[utf8]{inputenc}
\usepackage{times}
\usepackage{wasysym}
\usepackage{caption}
\usepackage{subcaption}
\usepackage{natbib}
\usepackage{multirow}
\let\oldtabular\tabular
\renewcommand{\tabular}{\scriptsize\oldtabular}
\usepackage[normalem]{ulem}

\begin{document}
\title{Transiting Planets with LSST. II. Period Detection of
Planets Orbiting 1~M$_{\odot}$ Hosts}
\author{Savannah Jacklin\altaffilmark{1,2}, 
Michael B. Lund\altaffilmark{2}, 
Joshua Pepper\altaffilmark{3}, 
Keivan G. Stassun\altaffilmark{2,4}}
\altaffiltext{1}{Department of Astrophysics and Planetary Science, Villanova University, Villanova, PA 19085, USA}
\altaffiltext{2}{Department of Physics and Astronomy, Vanderbilt University, Nashville, TN 37235, USA}
\altaffiltext{3}{Department of Physics, Lehigh University, Bethlehem, PA 18015, USA}
\altaffiltext{4}{Department of Physics, Fisk University, Nashville, TN 37208, USA}

\begin{abstract}
The Large Synoptic Survey Telescope (LSST) will photometrically monitor $\sim10^{9}$ stars for ten years.  The resulting light curves can be used to detect transiting exoplanets.  In particular, as demonstrated by \citet{Lund2014}, LSST will probe stellar populations currently undersampled in most exoplanet transit surveys, including out to extragalactic distances.  In this paper we test the efficiency of the box-fitting least-squares (BLS) algorithm for accurately recovering the periods of transiting exoplanets using simulated LSST data.  We model planets with a range of radii orbiting a solar-mass star at a distance of 7 kpc, with orbital periods ranging from 0.5 to 20 d.  We find that standard-cadence LSST observations will be able to reliably recover the periods of Hot Jupiters with periods shorter than $\sim$3~d, however it will remain a challenge to confidently distinguish these transiting planets from false positives. At the same time, we find that the LSST deep drilling cadence is extremely powerful: the BLS algorithm successfully recovers at least 30\% of sub-Saturn-size exoplanets with orbital periods as long as 20~d, and a simple BLS power criterion robustly distinguishes  $\sim$98\% of these from photometric (i.e. statistical) false positives.
\end{abstract}

\section{Introduction}

Starting in $\sim$2020, the Large Synoptic Survey Telescope (LSST) will begin operations, addressing topics including dark matter, dark energy, near-earth-objects (NEOs), variable stars, and many others.  Using the visual and infrared filters $ugrizy$, LSST will observe a 20,000 square degree area of the southern sky in a deep-wide-fast synoptic survey \citep{LSSTScience2009} to a depth of 24.5 mag per visit.  The LSST survey will produce light curves for $\sim10^{9}$ stars in the Milky Way and the Magellanic Clouds \citep{LSSTScience2009}. 

The first paper in this series, \citet{Lund2014}, found that LSST standard observations will be capable of detecting transiting planets in systems such as a $10 R_{\oplus}$ planet orbiting a G-dwarf, a $4 R_{\oplus}$ planet orbiting a K-dwarf, and a $2 R_{\oplus}$ planet orbiting an M-dwarf.  In this paper, we extend the analysis of \citet{Lund2014} to specifically quantify the recoverability of certain types of transiting planets as a function of planetary orbital period and planet radius.

\begin{figure}[!htb]
\centering
\includegraphics[width=.6\linewidth]{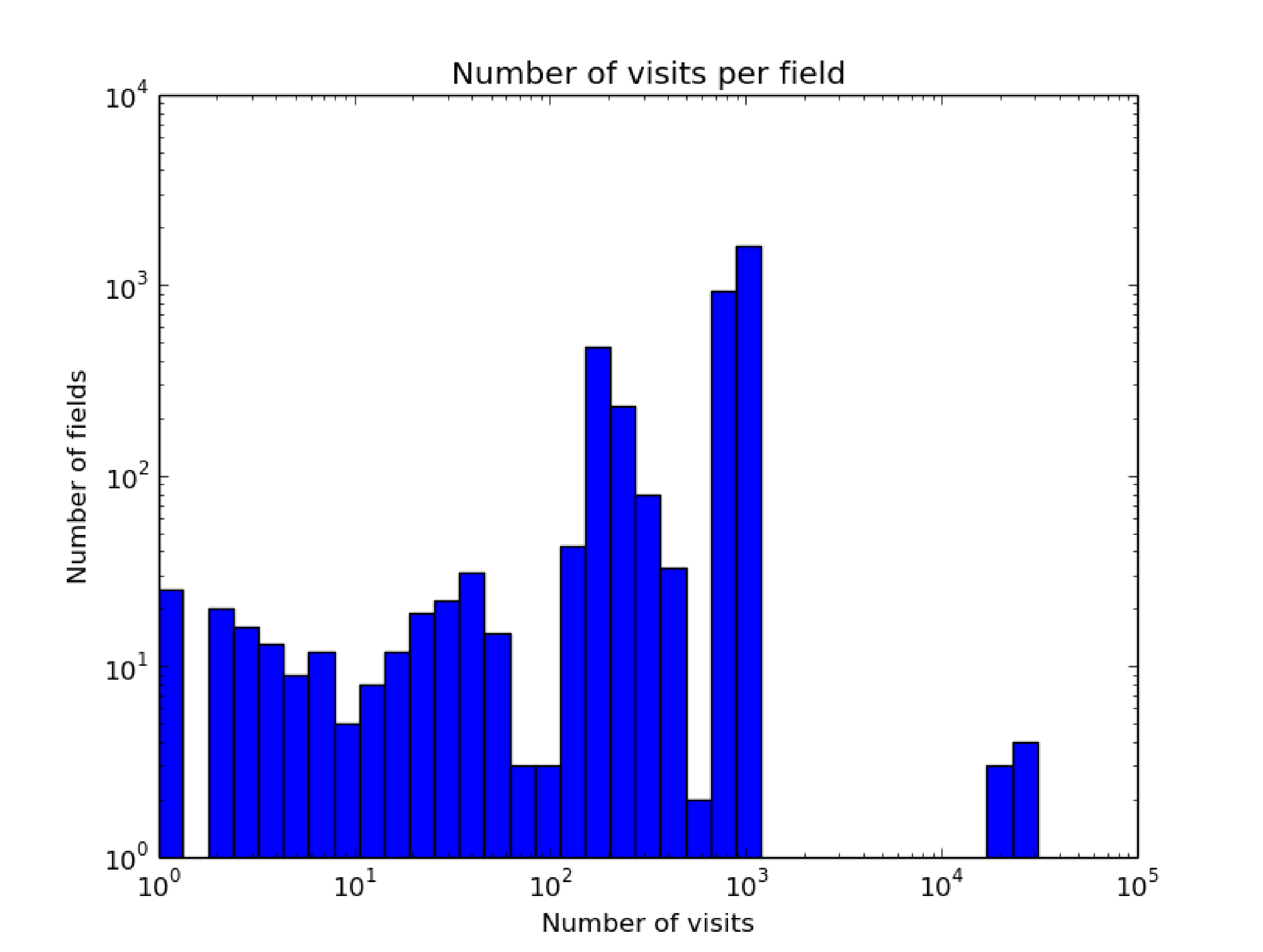}
\caption{Distribution of number of visits per field for a single LSST operations simulation. Deep-drilling fields show as a peak at $\sim$ 25,000 visits, and most regular cadence fields are in a peak at $\sim$ 1000 visits.}
\label{fig:visits}
\end{figure}

To assess the question of planet detection with LSST, it is particularly important to account for the limitations and complexity of the two cadence strategies that LSST will employ over the course of its ten-year expected lifetime \citep{Ivezic2008}.  LSST will acquire pairs of 15-s exposures, hereafter treated and referred to as 34-s visits.  Observations in most fields will have an average cadence of approximately 0.3~d$^{-1}$, but will usually involve multiple exposures in a single night separated by long gaps of many nights.  Each standard field will have on the order of $\sim1000$ distinct visits at regular cadence; see Fig \ref{fig:visits} for the distribution of visits per field.  Regular cadence observations will consume about 90\% of LSST's total observational schedule.  

In addition to the regular cadence, the LSST will observe a small number of fields ($\sim10$) at a much higher cadence, making about 10,000 visits over the course of the survey.  Deep drilling visits will utilize the remaining 10\% of LSST's observational schedule.  The sheer size of the observational area, the very large number of target stars, and the large number of observations comprising the light curves suggest that LSST has the potential for the discovery of transiting exoplanets, particularly in 
stellar populations that are faint, distant, or otherwise less thoroughly probed by previous transit surveys.  Although the frequency of observation is low compared to what is typically used in dedicated planet transit surveys (e.g., SuperWASP \citep{Doyle2012}, \emph{Kepler} \citep{Batalha2012}, HATNet \citep{Bakos2011}, KELT \citep{Pepper2007}, etc.), the total number of observed targets should drastically increase the statistical likelihood of observing a planetary transit.

This paper considers the impact of these different cadences on planet recoverability.  It is now known that there are a large variety of exoplanet system architectures, and therefore several astrophysical factors that ultimately influence the overall detectability of planets. These include host star size and rotational velocity; planet size, mass, and radius; planetary orbital period, eccentricity, and inclination; and host star photometric and spectroscopic variability.  We examine the recoverability of a transiting planet's periodic photometric signal, focusing specifically on large planets with sizes of 5--14 $R_{\oplus}$ orbiting a 1 M$_\odot$ star at 7 kpc from the Earth.

In Section~\ref{sec:methods}, we discuss the technical capabilities of LSST and the simulation of light curves using the LSST Operation Simulation (OPSIM), and describe our transit modeling and recovery methods and our photometric false positive analysis.  Section~\ref{sec:results} reviews the specific results of our research.  For each of our case studies, we report the planet period recoverability fraction at both regular and deep drilling cadence.  Section \ref{sec:discussion} discusses the future implications of this work and the next steps to be explored.  Section~\ref{sec:summary} briefly summarizes our conclusions.

\section{Methods\label{sec:methods}}

Our general approach is to generate a large number of synthetic light curves that simulate the expected LSST light curves in terms of both cadence and noise.  We then inject simulated planet transits into the light curves, and use standard period-finding methods to study the recoverability of the planet transit signals.

\subsection{Simulated LSST Light Curves}

We generate light curves using the methods described in \citet{Lund2014} which we briefly summarize here.  In general, light curves are created by first selecting the properties of the star and planet.  We select a stellar mass, and specify that the star is a dwarf.  The spectral type is determined by interpolating between masses in Table 15.8 in \citet{Cox2000}, and we determine the absolute magnitude of the star in all LSST filters based on that spectral type from \citet{Covey2007} and \citet{Hodgkin2009}.  The stellar radius is determined using the mass-radius relationship given in \citet{Beatty2008}.  For the specific cases studied in this paper, we adopt 1 M$_\odot$ and thus 1 R$_\odot$ in all cases for the host star mass and radius, and we adopt a distance of 7 kpc in all cases for the stellar distance.

Next, the planet radius and period are selected.  Once those properties are assembled, noiseless and continuous light curves are generated by treating the transits as simple boxcars.  The transit depth is based on the ratio of star and planet radii, and the duration is based on the planet orbital period and stellar radius.  All simulated systems assume equatorial transits and circular orbits.

\begin{figure}[!htb]
\centering
\includegraphics[width=.55\linewidth]{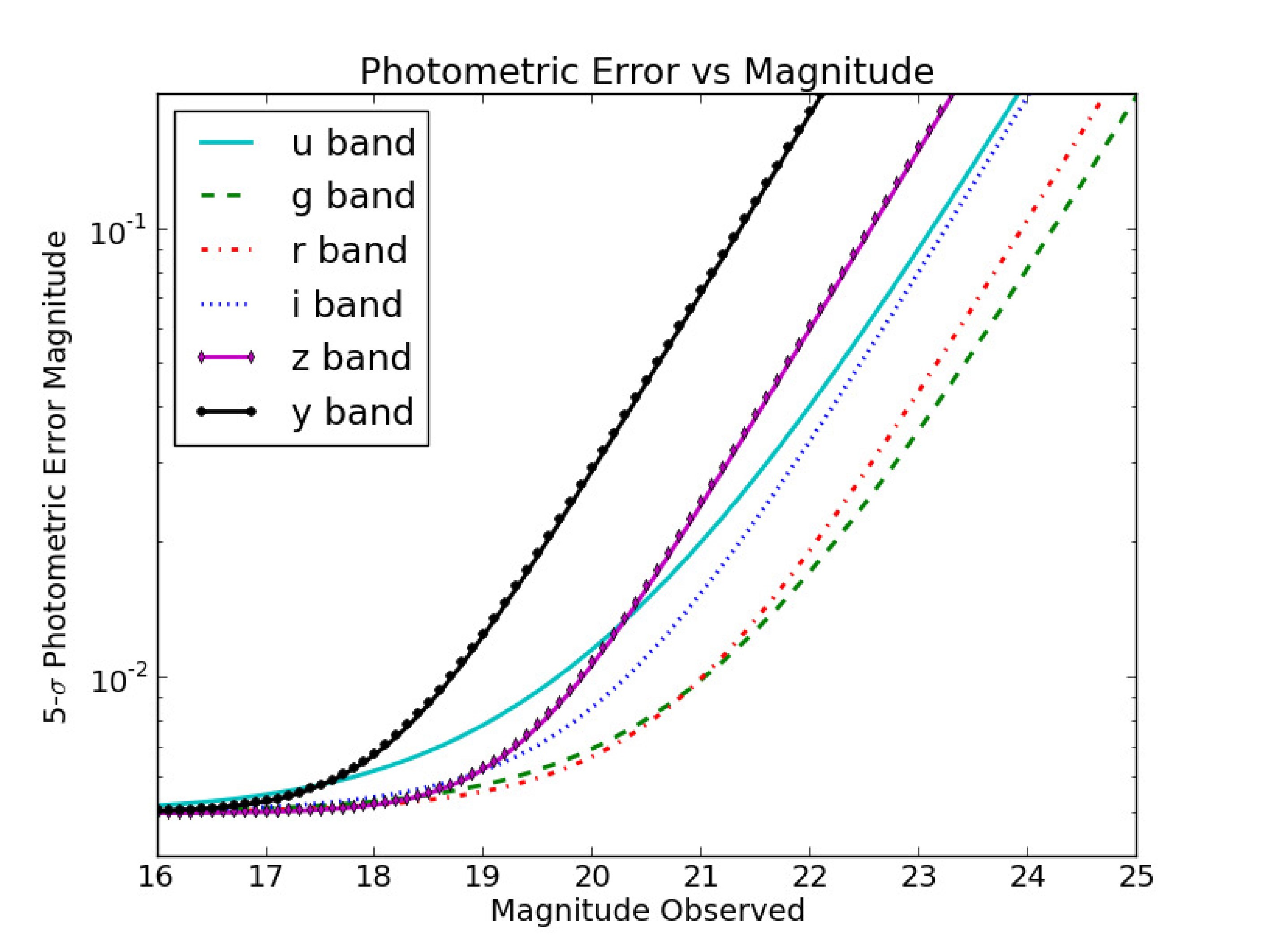}
\caption{Per-visit LSST photometric error as a function of apparent magnitude for each filter.}
\label{fig:noise}
\end{figure}

The light curves are then sampled in time using the operation simulation (OPSIM) v2.3.2, run 3.61 results released by LSST\footnote{Available at \url{https://www.lsstcorp.org/?q=opsim/home}}.  The OPSIM results represent ten years of scheduled observations for LSST, accounting for downtime due to weather and other factors. We then place the system at a given distance from the Earth, and determine the apparent magnitude of the host star for each of the LSST filters ($ugrizy$).  Based on the stellar apparent magnitude we compute the expected per-visit photometric precision of the LSST observations of the star for each filter using the equation \citep{Ivezic2008}:
\begin{equation}
\sigma_{tot}^2 = \sigma_{sys}^2 + \sigma_{rand}^2
\end{equation}
where $\sigma_{sys}$ is an assumed systematic noise floor and is set to $0.005$ mag for all filters.  The random photometric noise $\sigma_{rand}^2$ varies by observational filter, and is characterized by a parameter $\gamma$, unique to each filter.  To compute the random noise per LSST visit at the zenith, the noise model is given by the equation:
\begin{equation}
\sigma_{rand}^2 = (0.04 - \gamma)x + \gamma x^2 
\end{equation}
where $x = 10^{(m-m_{5})}$ and $m_{5}$ is the 5$\sigma$ limiting magnitude for a given filter, which depends on the local observing conditions. Since $m_5$ is different for each visit, it is generated by the OPSIM. The values of $m_{5}$ and $\gamma$ are described by \citet{Ivezic2008}.  The total photometric noise as a function of apparent magnitude in each filter is shown in Figure \ref{fig:noise}.

The random noise tends to be largest in the $u$, $z$, and $y$ filters, and main sequence dwarf stars tend to be intrinsically much fainter in those filters as well.  We therefore discard the visits using the $u$, $z$, and $y$ filters, since their larger noise makes them less useful for transit detection.  In the final step we median-subtract each of the remaining light curves (those in the $g$, $r$ and $i$ filters) and combine them into one master light curve.

\subsection{Recovery of Transits, Transit Periods and Photometric False Positives\label{sec:pertest}}

The principal metric in this study is the recoverability of the period of the simulated planet transits. For this purpose, we utilize the standard BLS algorithm \citep{Kovacs2002} as implemented in the VARTOOLS software tool \citep{Hartman2008}.

We first examine how well BLS is able to recover a sample transiting planet.  First, we generate 1000 simulated light curves of a typical Hot Jupiter.  This system consists of a solar-type star (1 M$_\odot$, 1 R$_\odot$) with a $10 R_\oplus$ planet in a 4.2-d orbit.  We place the star at 7 kpc from the Earth, in order to situate it at a distance where the apparent magnitude of the star is near the bright end of the LSST magnitude regime.  The 1000 simulated light curves represent 1000 realizations of the LSST noise model and in each we randomize the orbital phase of the planet transit.

\begin{figure}[!htb]
\centering
\begin{subfigure}{.4\textwidth}
  \centering
  \includegraphics[width=\linewidth]{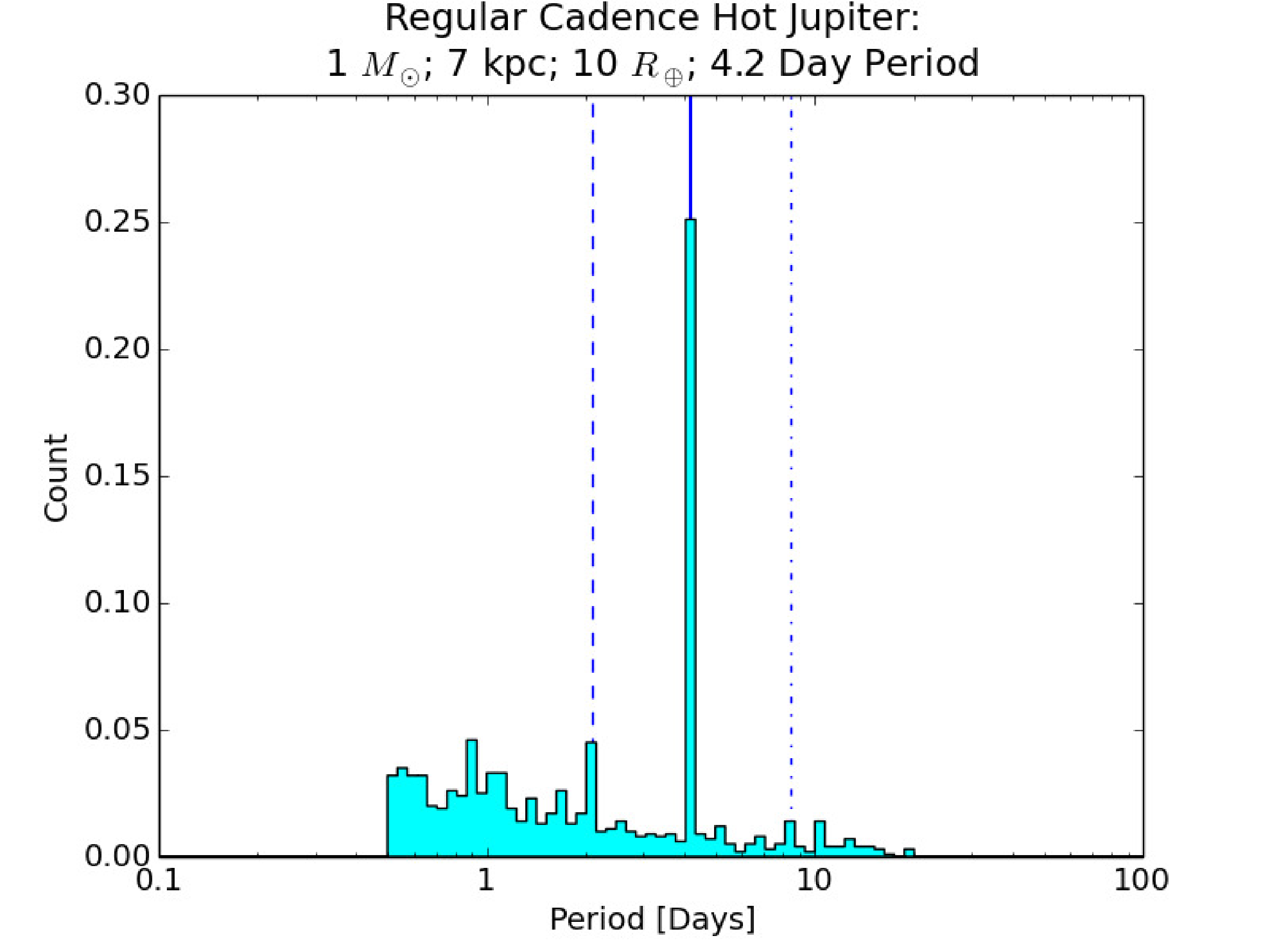}
\end{subfigure}
\begin{subfigure}{.4\textwidth}
  \centering
  \includegraphics[width=\linewidth]{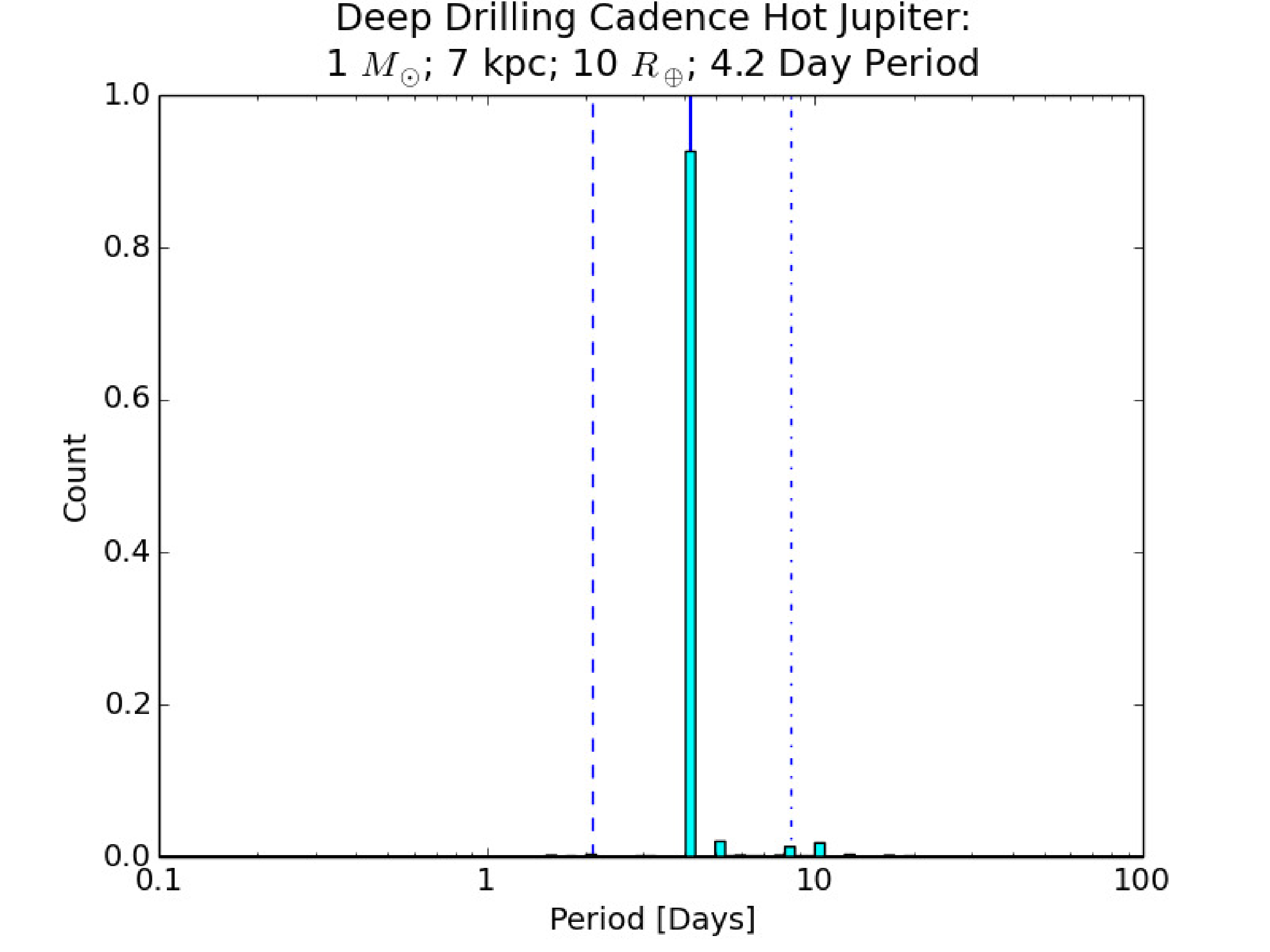}
\end{subfigure}
\caption{Normalized histograms depicting the top BLS recovered period for a $10 R_\oplus$ hot Jupiter orbiting a one solar mass star at 7 kpc from Earth with a 4.2 day input period, for regular cadence ({\it left}) and deep drilling cadence ({\it right}).  The bold vertical line marks the input period, and the two dashed lines on either side mark the half and double period.}
\label{fig:HJ}
\end{figure}

We run BLS on each light curve and compare the top period returned by BLS (as measured by the BLS signal to noise ratio) to the input period.  Figure \ref{fig:HJ} shows a histogram of the resulting periods for the 1000 realizations, using light curves with regular LSST cadence (\textit{a}) and the deep drilling cadence (\textit{b}).  For this initial test case, we see that BLS recovers the correct input period with an accuracy of 0.1\% in at least 25\% of cases for regular cadence and at least 91\% of cases for deep drilling cadence. 

Aside from the question of recovering the transit periods correctly, an important issue for LSST will be identifying the transits with some measure of statistical confidence over various types of false positives.  For many transit surveys, astrophysical false positives are most important to eliminate \citep{Brown2003}.  Those will be extremely important for LSST transit searching, but in this paper, we focus only on photometric false positives, where the photometric noise randomly produces transit-like features in light curves.  Assuming Gaussian  photometric noise, we can quantify how well the BLS power can distinguish a real input transit signal from a photometric false positive at a given confidence level.

\begin{figure}[!htb]
\centering
\begin{subfigure}{.45\textwidth}
  \centering
  \includegraphics[width=\linewidth]{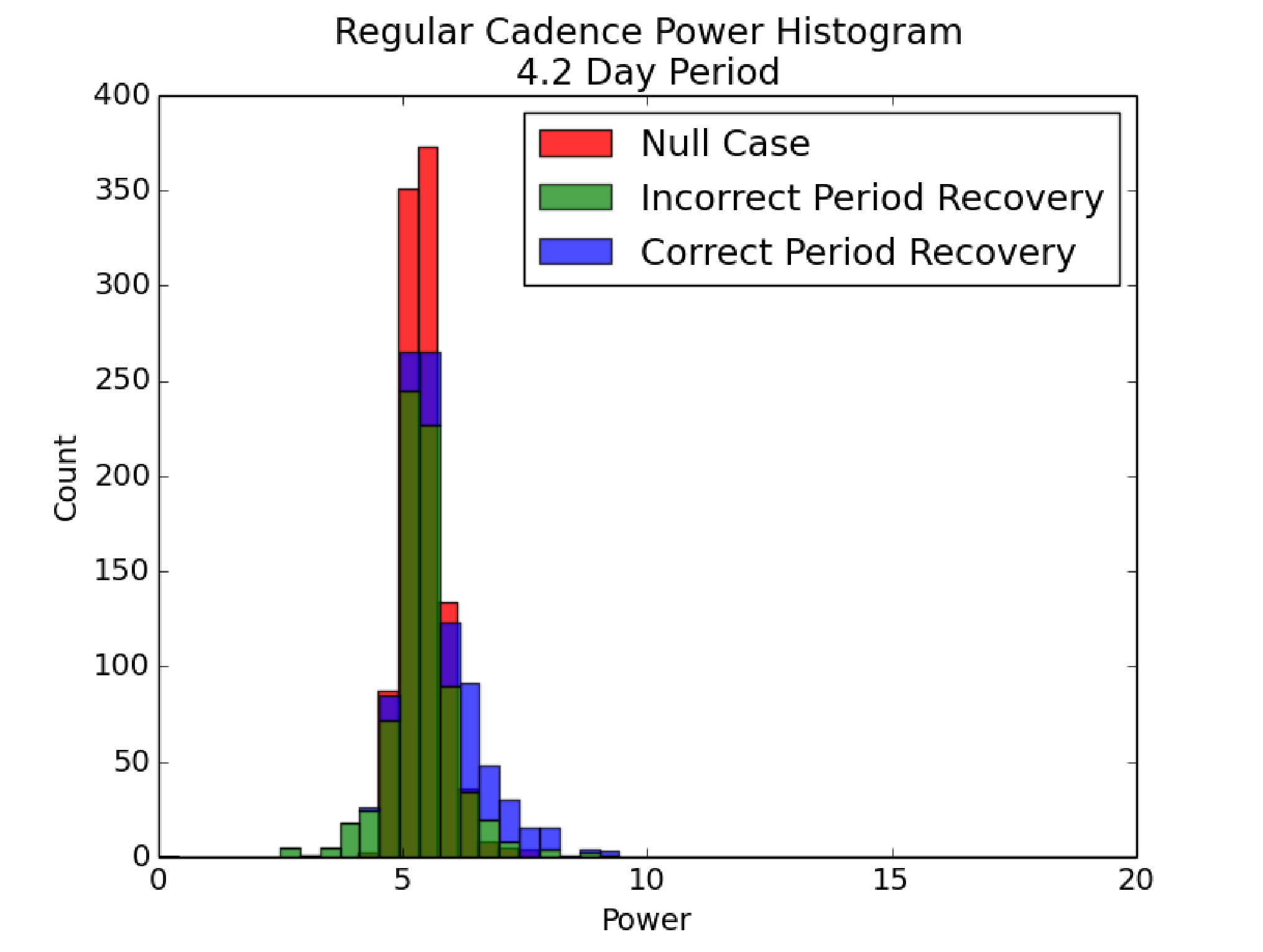}
  \caption{Regular Cadence}
  \label{fig:top_power_reg}
\end{subfigure}%
\begin{subfigure}{.45\textwidth}
  \centering
  \includegraphics[width=\linewidth]{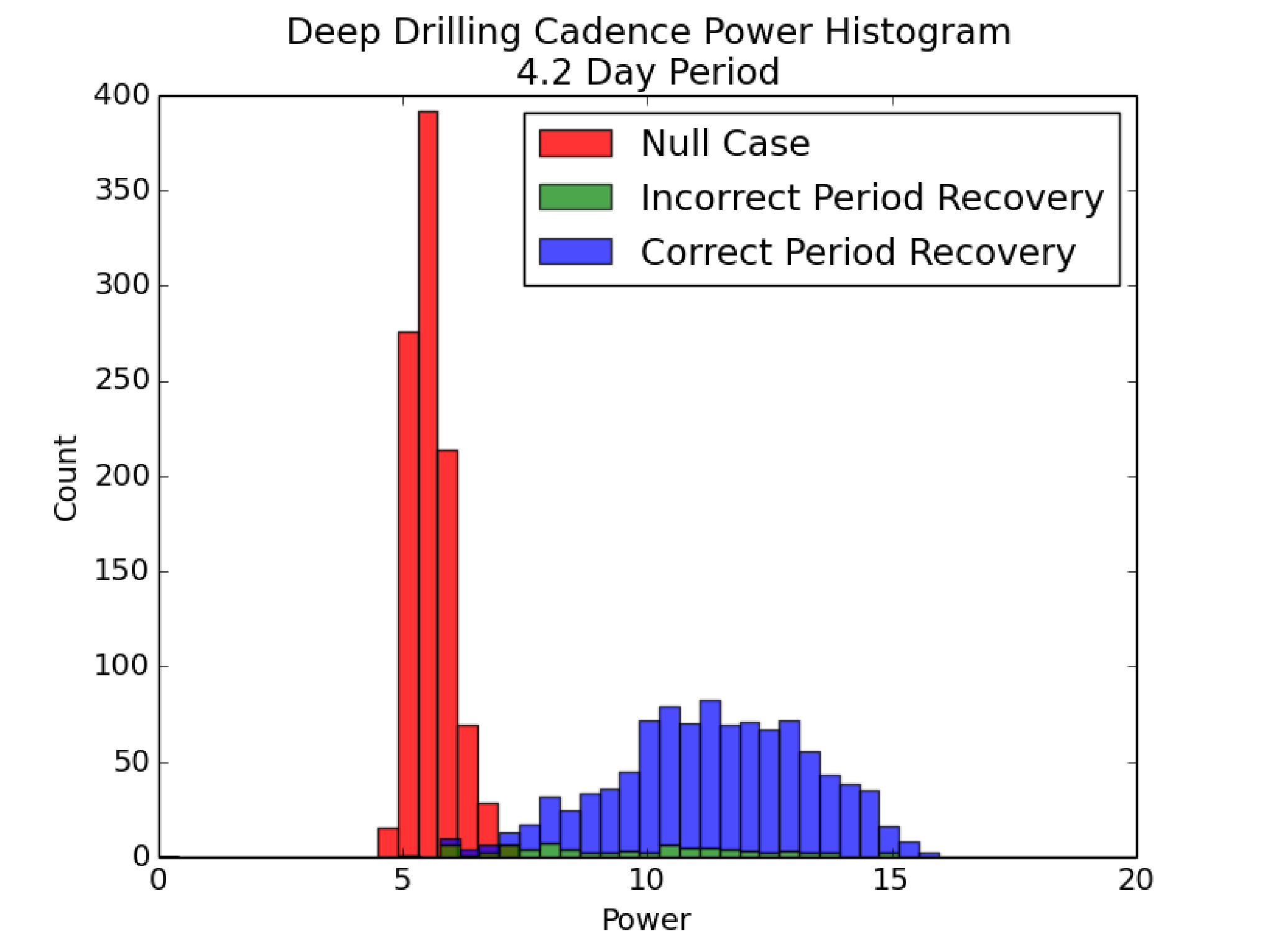}
  \caption{Deep Drilling Cadence}
  \label{fig:top_power_dd}
\end{subfigure}
\caption{Histograms depicting the power for the top BLS recovered period for a $10 R_\oplus$ Hot Jupiter orbiting a one solar mass star at 7 kpc from Earth with a 4.2 day input period, for regular cadence ({\it left}) and deep drilling cadence ({\it right}) with respect to the null case shown in red. Green represents cases where the correct period was not recovered, and blue represents cases where the correct period was recovered.}
\label{fig:power}
\end{figure}

Figure~\ref{fig:power} examines the distributions of BLS power for simulated transits (blue and green histograms) versus the ``null" case of light curves without any injected transits (red histogram). Because our simulations involve 1000 simulated light curves, the maximum BLS power in the 1000 null light curves in essence represents the power above which we can expect a photometric false-positive probability of $<$0.1\%.  In the case of a regular cadence field (left panel), we find that light curves with correctly recovered periods have, on average, higher BLS power than light curves for which the correct period is not recovered or for the null case, as expected.  However, only 7\% of all regular cadence light curves have correct periods recovered with a BLS power above the 99.9\% confidence threshold defined by the null case. Clearly, photometric false positives will be a serious concern for regular cadence fields, at least using the BLS criteria we have examined here.

The situation is much improved, however, in the deep drilling cadence. Here, 98\% of light curves with correctly recovered periods having BLS power above the 99.9\% confidence threshold. From this simple analysis we see that the deep drilling fields not only recover the correct period at a higher rate, but also that a much larger proportion would survive a very simple cut on BLS power to exclude photometric false positives. 

For the remainder of this paper we focus exclusively on the assessment of transit {\it period} recoverability, though it remains an important future topic of investigation whether there are more sophisticated algorithms that would be useful in identifying true transits in both regular and deep drilling cadence data, such that the good period recoverability potential that we find below is fully harnessed with those data.

\section{Results\label{sec:results}}
We now apply our period recoverability test as in Sec.~\ref{sec:pertest} to a broad range of planet radii and orbital periods.  First, we examine what fraction of planets are correctly recovered across a range of input periods for several types of planets.  Second, we broaden that analysis to look at transit recovery across a range of period and radius parameter space.  In all cases, we define a correctly recovered transit as one in which the period of the strongest BLS signal matches the input period to a 0.1\% accuracy.

We simulate six different planets sizes, representing "Hot Jupiters" with radii of 12~R$_\oplus$, 11~R$_\oplus$, 10~R$_\oplus$, "Hot Saturns" with radii of 9~R$_\oplus$ and 8~R$_\oplus$, and a "Hot Neptune" with a radius of 4~R$_\oplus$.  In each case, we simulate the same host star representing a Sun-like dwarf with mass of 1~M$_\odot$ and radius of 1~R$_\odot$ at a distance of 7 kpc, and in each case we vary the planet orbital period over the range 0.5--20~d in increments of 0.1~d.  For each of the exemplar cases, we create 1000 light curves for each input period, for both regular cadence and deep drilling cadence, as in Section \ref{sec:methods}. 

\begin{figure}[!htb]
\centering
\begin{minipage}[b][0.63\textheight][s]{0.8\linewidth}
  \centering
  \includegraphics[height=0.3\textheight,width=\linewidth]{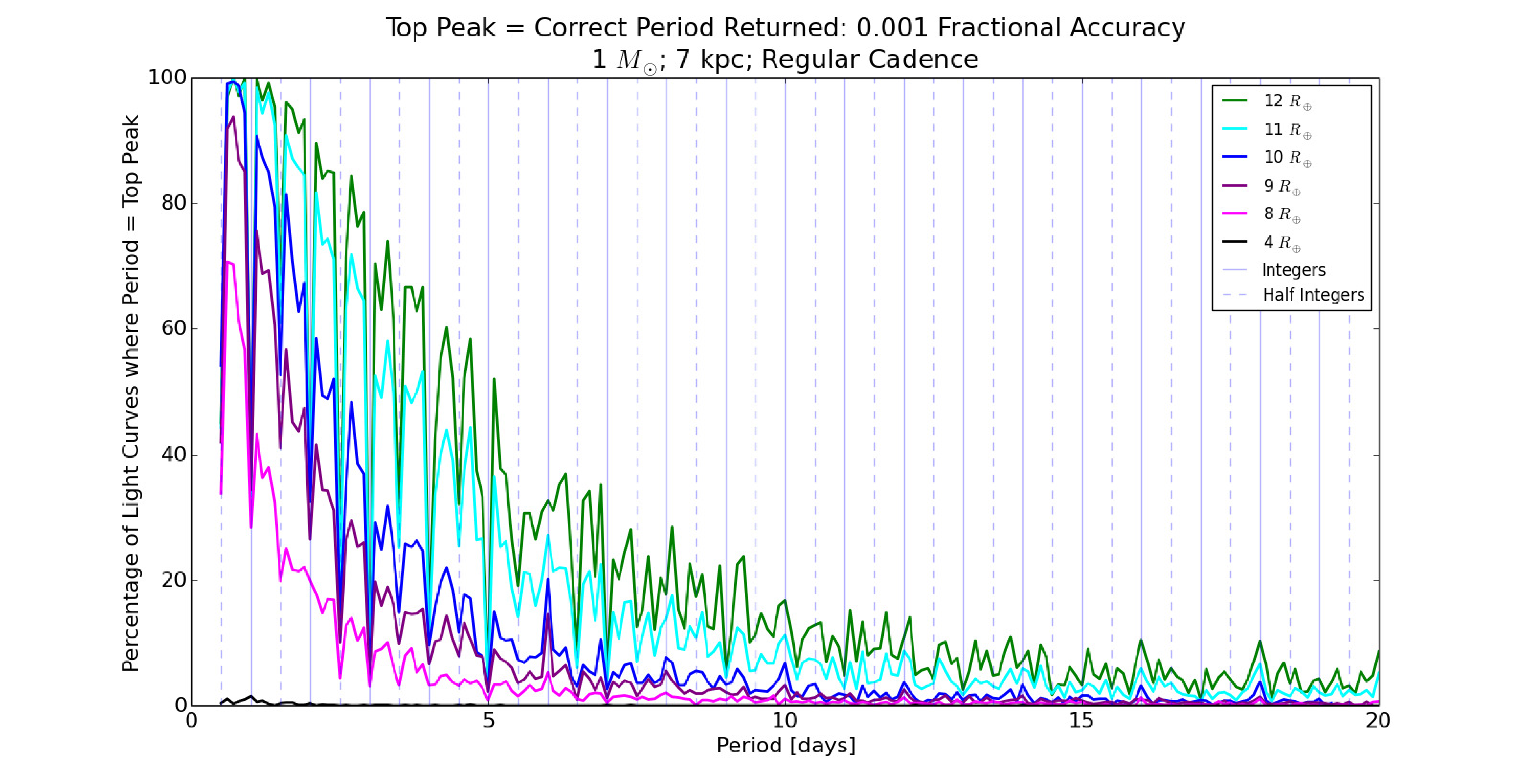}
  \includegraphics[height=0.3\textheight,width=\linewidth]{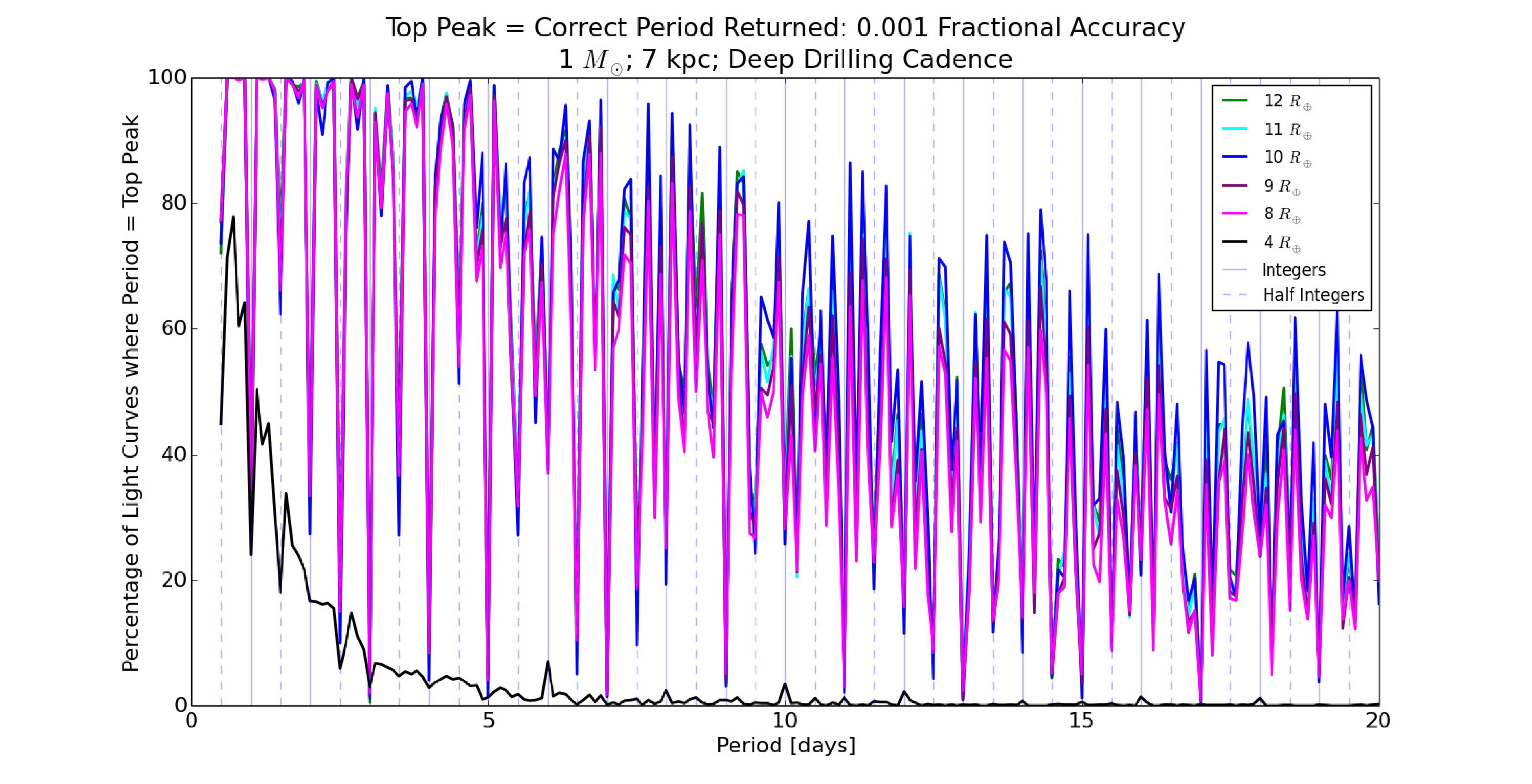}
\end{minipage}
\caption{Regular (\textit{top}) and deep drilling (\textit{bottom}) cadence period recoverability of 4, 8, 9, 10, 11, 12 $R_{\oplus}$ transiting exoplanets about a 1 $M_{\odot}$ star at 7 kpc.  The larger the planetary radius and the shorter the period, the more likely it is that BLS will recover the top period.  Period recoverability is much more accurate at deep drilling cadence than at regular cadence, with significant recoverability exhibited for most tested planetary radii out to 20~d periods and beyond.}
\label{fig:THJ}
\end{figure}

\begin{figure}[!htb]
\centering
\begin{subfigure}{.49\textwidth}
  \centering
  \includegraphics[width=\linewidth]{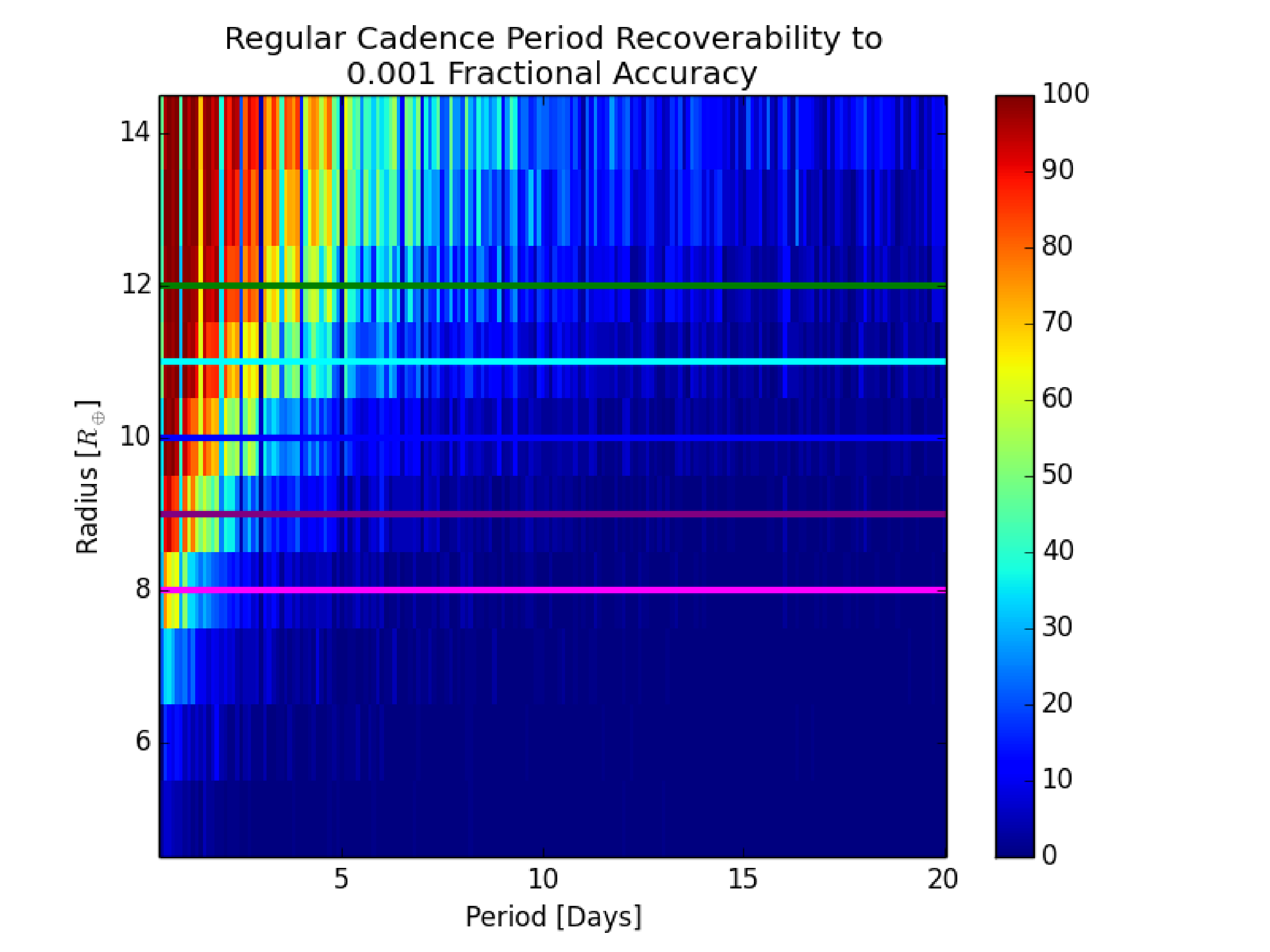}
  \label{fig:top_hist_reg}
\end{subfigure}%
\begin{subfigure}{.49\textwidth}
  \centering
  \includegraphics[width=\linewidth]{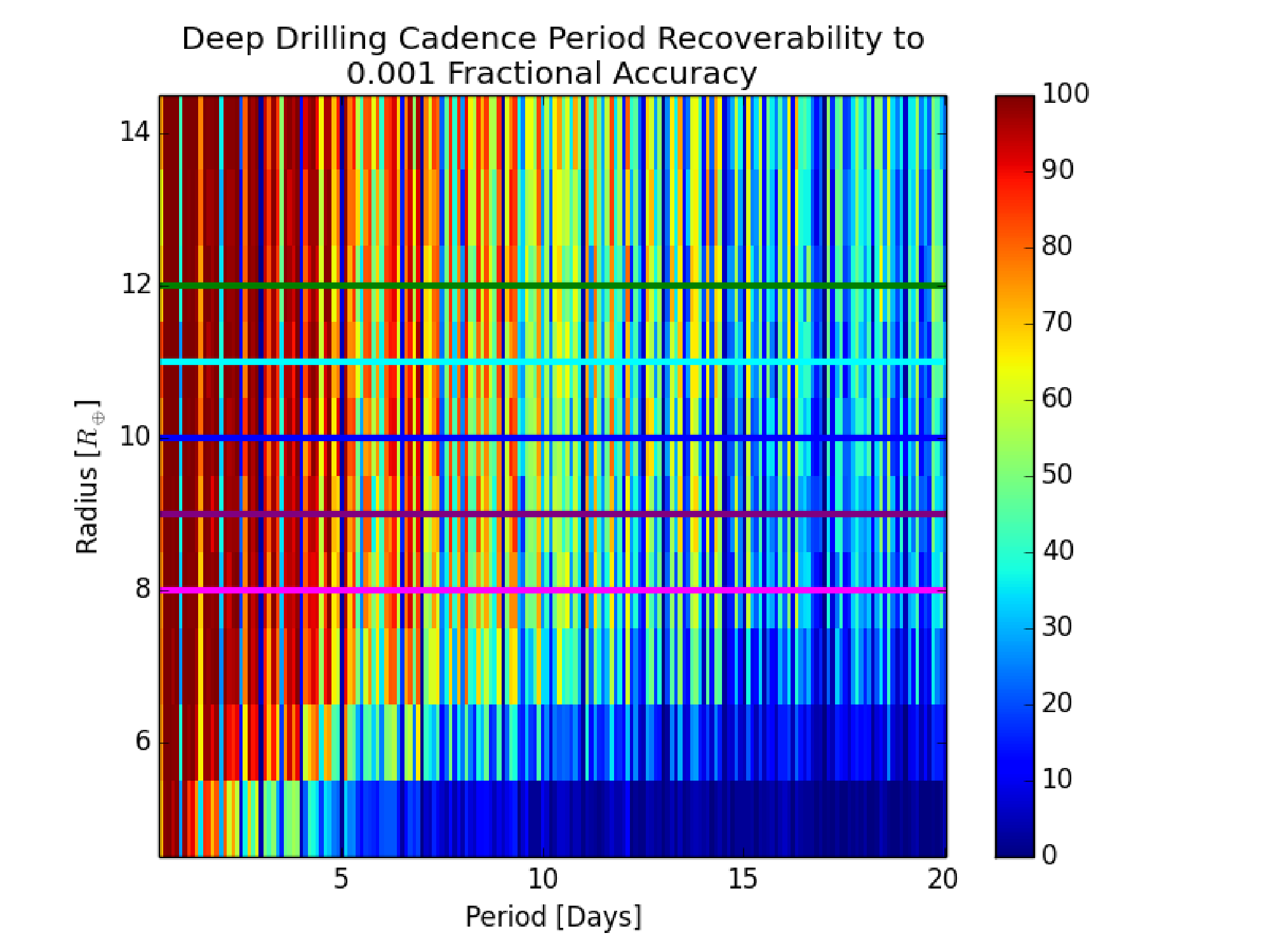}
  \label{fig:top_hist_dd}
\end{subfigure}%
\label{fig:perrad}

\centering
\begin{subfigure}{.49\textwidth}
  \centering
  \includegraphics[width=\linewidth]{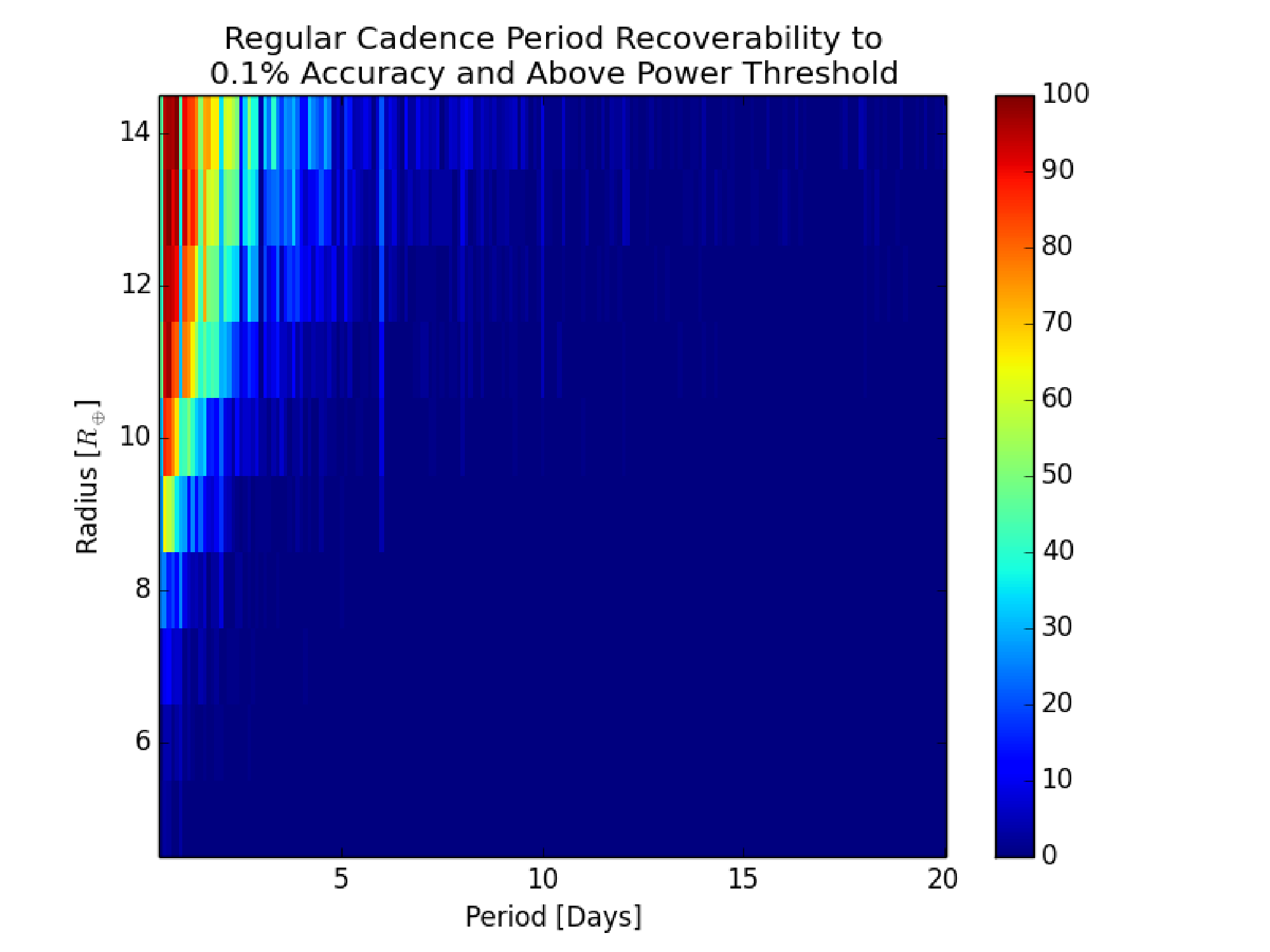}
  \caption{Regular Cadence}
  \label{fig:perpow_reg}
\end{subfigure}%
\begin{subfigure}{.49\textwidth}
  \centering
  \includegraphics[width=\linewidth]{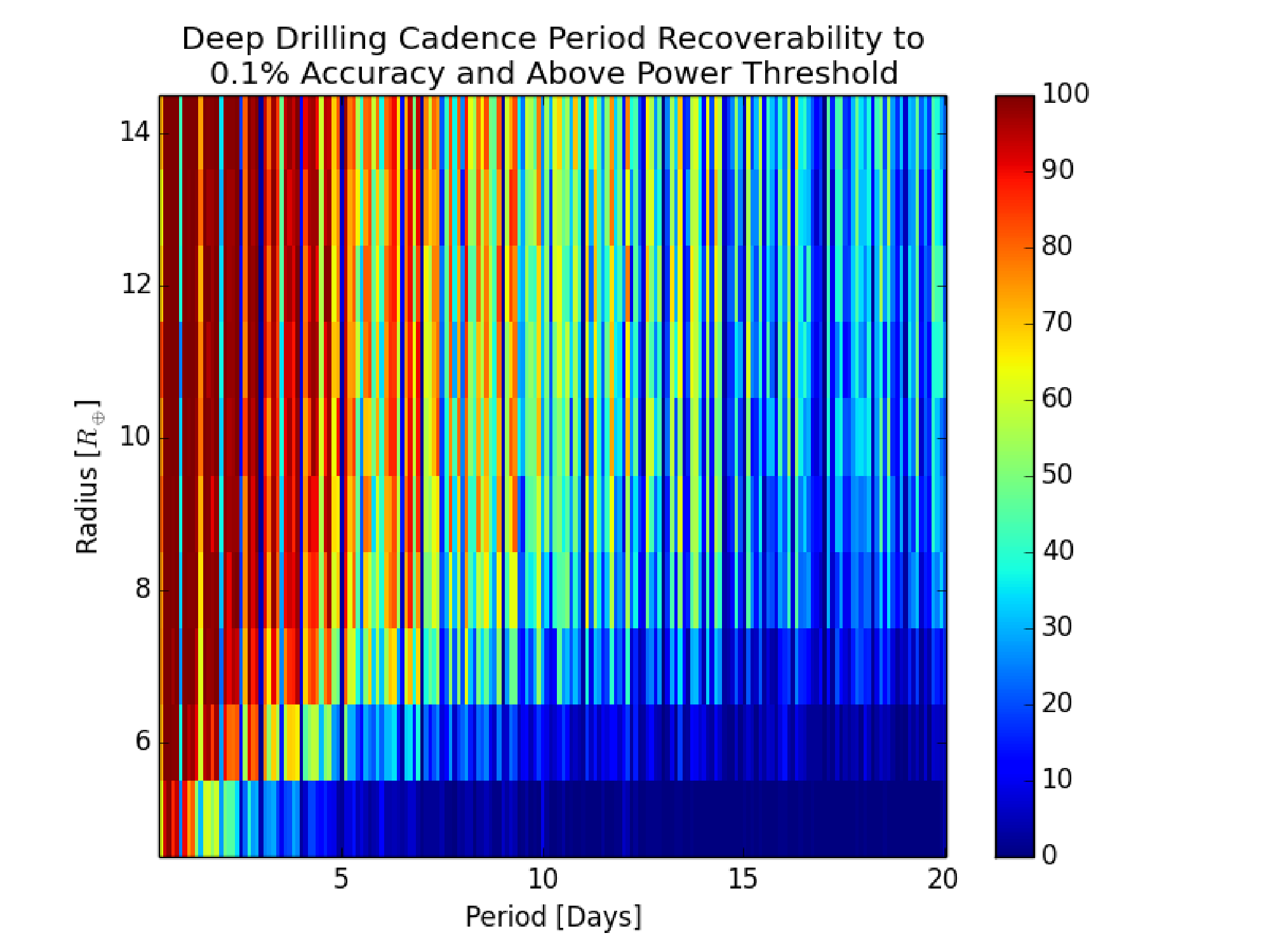}
  \caption{Deep Drilling Cadence}
  \label{fig:perpow_dd}
\end{subfigure}%
\caption{Two-dimensional histograms representing period recoverability across period and radius space.  The top two plots represent period recoverability considering only top BLS peak accuracy.  The bottom two plots include an additional power threshold with all detections also having a BLS power greater than 7.69526 for regular cadence and 7.32893 for deep drilling cadence.  Warm colors indicate high recoverability and cool colors indicate lower percentage recoverability as is shown by the color bar.  Deep drilling cadence (\textit{right}) exhibits considerably higher recoverability in both period and radius space than regular cadence (\textit{left}). Colored horizontal lines designate radii which are represented in Fig.~\ref{fig:THJ}.}
\label{fig:PHJ}
\end{figure}

The period recoverability results for the specific planet radii are shown 
in Figure \ref{fig:THJ}, for both regular and deep drilling cadence, using vertical lines to emphasize detail at the integer and half-integer day periods.  We also represent the results of period recoverability across period and radius space using two-dimensional histograms to convey the broader trends, as shown in Figure \ref{fig:PHJ} (\textit{a}) at regular cadence and Figure \ref{fig:PHJ} (\textit{b}) at deep drilling cadence. Warm colors such as red and orange indicate high fractional accuracy period recoverability and cool colors such as blue and green indicate lower fractional accuracy. We use the maximum power from the null cases displayed in Figure~\ref{fig:power} to establish a power threshold that represents detections with a photometric false positive rate $<$ 0.1\% using our current BLS detection algorithm. Irrespective of the power cut, we find that the deep-drilling fields exhibit substantially higher period recoverability rates, and with the power cut implemented there is only a small decrease in recoverability.  This indicates that the current algorithms we are using would be sufficient to identify candidate transiting planet light curves. For regular fields, we find that while there is still a sizeable number of light curves that return the correct period, the power is often too low for us to identify them as candidates and we are only left with significant number of detections for very large planets with very short periods. There will be a very large number of light curves that the regular fields will contain, and any improvements in candidate selection using a more sophisticated algorithm to identify planet candidates will result in an increase in planet yield. These improvements will also likely include increasing the parameter space that is being reliably explored by extending it to longer periods and smaller radii.  Examining how best to improve our current algorithm will be an interesting question for future work.

As expected, the results are similar to previous calculations of transit recoverability \citep{VonBraun2009, Burke2006}, with strong biases towards greater recoverability  for large planets and short periods.  The larger number of observations with deep drilling cadence significantly increase the recovery rates.  In particular, in the standard cadence the recoverability is strongly affected by planetary radius.  Inflated Hot Jupiters ($R>11R_{\oplus}$) exhibit recoverability greater than 50\% out to periods of $\sim$4 days. Non-inflated Hot Jupiters and Saturns are recoverable out to $\sim$2 days. Hot Neptunes are essentially undetected, with a maximum recoverability of $\sim 2\%$ at a 1-d period.  The deep drilling fields are quite different. Large giant planets ($R>7R_{\oplus}$) are recoverable at a rate of greater than 60\% out to periods of 10 days, and at rates from 30\% to 50\% out to at least 20 days.  For small planet sizes of  $R\approx6R_{\oplus}$, the recoverability drops quickly, and only short period ($P<5$ d) planets are recovered.

In all cases, we also see that the period sensitivity is strongly influenced by the diurnal sampling, showing sharp declines in sensitivity at integer and half-integer day periods, which is not surprising.  Interestingly, however, while the window function results in drops in period recoverability at integer and half-integer periods when the average period recoverability is high, there are actually increases in recoverability at the integer and half-integer periods when the average period recoverability is very low (e.g., black curve in Figure \ref{fig:THJ}).  This is because there is an increase in the number of observed transits for some planets when the period is at a half- or full-day integer, such that transits all occur at night when LSST can observe.

\begin{figure}[!htb]
\centering
\includegraphics[width=0.8\linewidth]{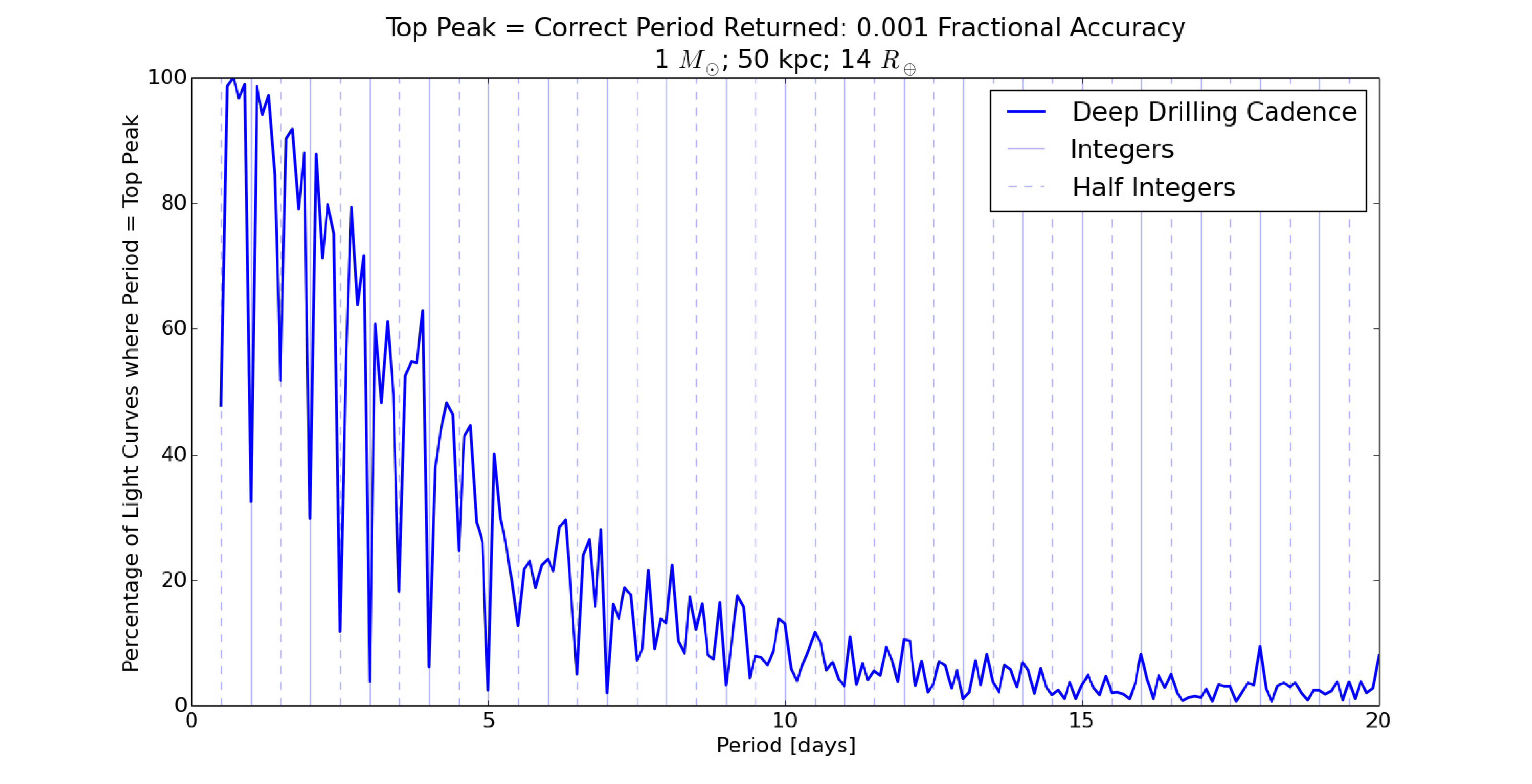}
\caption{Deep drilling period recoverability for a 14 $R_{\oplus}$ Hot Jupiter about a 1 $M_{\odot}$ star at 50 kpc.  This simulation exhibits LSST's period recoverability potential in the Large Magellanic Cloud.}
\label{fig:lmcdd}
\end{figure}

Finally, we analyze deep drilling period recoverability of a planet in the Large Magellanic Cloud. We simulated an inflated Hot Jupiter with a radius of 14 $R_{\oplus}$ around a $1M_{\odot}$ star at 50 kpc, and we use only the $g$ filter which has the lowest photometric noise (Fig.~\ref{fig:noise}).  The resulting recoverability is shown in Fig.~\ref{fig:lmcdd}, which clearly shows that extragalactic planet recoverability is possible using LSST for such a system.

\section{Discussion\label{sec:discussion}}

From this work, we find clear indications that LSST will be capable of period recoverabilty for many interesting types of transiting planets. For the majority of the survey carried out at the regular cadence, we have seen that it should be possible to correctly recover the period of $\sim$50\% of Hot Jupiter ($\sim 10 R_{\oplus}$) transits around solar-type stars at orbital periods of up to 3--4~d. While we have not explicitly explored other planet/star configurations here, this does demonstrate that LSST should be capable of recovering planets whose transits cause drops in brightness of $\sim$1\%, and thus we expect that smaller planets orbiting smaller stars should also be detectable (although presumably at closer distances than the 7 kpc distance of the G-dwarf simulated here).

The deep drilling fields, while a much smaller portion of the planned LSST survey, have much greater potential for planet detection. In those fields, we show that it should be possible to accurately recover Hot Jupiters $\sim$40\% of the time with periods as long as $\sim$20~d, the outer limit of our period search. We also find that in these deep drilling fields it should be possible to recover Hot Neptunes ($\sim 4 R_{\oplus}$) with orbital periods of 1--2~d. While Hot Jupiters have been the most well-studied population of exoplanets, surveys such as \emph{Kepler} have shown that Hot Neptunes are much more common, and thus represent a less well-studied population of exoplanets. The detection of smaller planets (particularly Hot Neptunes) using LSST is an exciting prospect, as they are generally smaller in size than what current ground-based searches have been able to discover around solar-type stars thus far.

Any detections that LSST makes will represent planet candidates -- not confirmed planets.  Due to the faintness of the stars that LSST will observe, traditional confirmation methods will likely not be feasible. In both this work and \citet{Lund2014}, we have only shown that periodic transiting events can be recovered from LSST light curves. While this does show that the signal from a planet transit should be recoverable, it does not yet address the ability to differentiate between transiting planets and astrophysical false positives (such as eclipsing binaries). In addition, as shown in Sec.\ 2.2, for the regular cadence data in particular it will be challenging to disambiguate many or most of the types of transits we have simulated from statistical false positives, despite the accuracy with which the transit periods can be recovered. Our goal in a future paper will be to explore ways to differentiate between true transiting planets and other phenomena, as well as to investigate more robust methods for confidently identifying true transits in the regular cadence data. Also, since follow-up observations which characterize LSST-detected planets in terms of atmospheric composition, thermal properties, or dynamics will generally not be an option, the main power of the detection of planet candidates will be to determine statistical populations.  Learning how to correct the LSST planet candidate population for contamination by other astrophysical signals will therefore be crucial for statistical exoplanetary census studies.

In future studies we will examine the detection of Hot Jupiters at even longer periods, stellar hosts of different types and distances, and specific stellar populations. In addition, both this paper and \citet{Lund2014} have focused on recoverability at the end of ten years of LSST operations. We intend to examine planet detection rates at different points during the ten years of observation.

\section{Summary\label{sec:summary}}
The LSST mission will generate approximately one billion stellar light curves over more than half the sky. The noise in these light curves will be sufficiently low that the signal of a transiting planet can be detected.  In this paper we have simulated the expected LSST observing cadences and light curve noise properties for the specific case of a solar-type star at 7000 pc distance, into which we injected simulated transits of planets of various sizes and orbital periods. We tested the ability of the standard BLS transit-period detection algorithm to correctly recover the input transits.

We find that there will be sufficient sensitivity to detect Hot Jupiters, Saturns, and Neptunes orbiting solar-type stars at distances of 7 kpc.  However, the LSST cadence matters greatly for period recoverability.  For regular fields, we find that Hot Jupiters are recovered from close to 100\% of the time at very short periods (less than 3~d) to $\sim$ 10\% of the time out to periods of around 10~d, while smaller planets (super Neptunes) are not recovered at any significant rate. However, in deep-drilling fields we find that Hot Jupiters can be recovered at rates of at least $\sim$ 40\% out to periods of 20~d, and sub-Saturn size planets (super Neptunes) are detectable at short periods (up to 2--3~d).  The LSST cadence also matters greatly from the standpoint of detecting these exoplanet transits with statistical confidence and ruling out false positives. In the BLS simulations studied here, we find that false positive period detections arising from pure chance due to photometric noise can be confidently ruled out in only $\lesssim$10\% of cases, whereas in the deep drilling cadence we find that photometric false positives can be ruled out in $\gtrsim$98\% of cases.

There is considerable room for further investigation at longer periods in the deep-drilling fields. LSST's ten-year observing baseline, and its recoverability of Hot Jupiters out to orbital periods of at least 20 d, indicate that LSST may be able to look for transiting planets in even longer periods, a regime where many transit searches lack coverage. We also have not yet addressed the ability of LSST to detect planets around other types of stars, or around stars that are at much greater distances. However, based upon the preliminary work here, we continue to demonstrate that LSST is poised to provide a unique contribution to the study of transiting exoplanets.

\acknowledgments
We thank LSST, Vanderbilt University, Lehigh University, and Villanova University, and acknowledge funding support from the NSF REU program at Vanderbilt University.

\newpage

\bibliographystyle{apalike}
\bibliography{libAAS}

\end{document}